\documentclass[useAMS,usenatbib,usegraphicx,uselandscape]{mn2e}
\usepackage{epsf}
\usepackage{psfig}
\usepackage{lscape}
\usepackage{times}
\usepackage{latexsym}
\usepackage{longtable}

\title[Stars with infrared excess]{Tycho 2 stars with infrared excess
in the MSX Point Source Catalogue}

\author[Clarke et al.]{A. J. Clarke\thanks{E-mail:
ajc@ast.leeds.ac.uk (AJC); roud@ast.leeds.ac.uk (RDO);
sll@ast.leeds.ac.uk (SLL)}, R. D. ~Oudmaijer\footnotemark[1] and
S. L. Lumsden\footnotemark[1]\\ School of Physics and Astronomy,
University of Leeds, Leeds, LS2 9JT, UK}

\begin{document}

\maketitle

\begin{abstract}
Stars of all evolutionary phases have been found to have excess
infrared emission due to the presence of circumstellar material. To
identify such stars, we have positionally correlated the infrared MSX
point source catalogue and the Tycho 2 optical catalogue. A near/mid
infrared colour criteria has been developed to select infrared excess
stars. The search yielded 1938 excess stars, over half (979) have
never previously been detected by IRAS. The excess stars were found to
be young objects such as Herbig Ae/Be and Be stars, and evolved
objects such as OH/IR and carbon stars. A number of B type excess
stars were also discovered whose infrared colours could not be readily
explained by known catalogued objects. 

\end{abstract}


\section{Introduction}
\label{intro}

The discovery of excess far-infrared emission from $\alpha$ Lyrae
by the Infrared Astronomical Satellite (IRAS) by \citet{AuLyrae}
demonstrated for the first time that proto-planetary material around
other stars could be both detected and studied. Further studies of
IRAS point sources revealed that many types of star displayed strong
emission at IRAS wavelengths. This excess is mostly due to thermally
re-radiating circumstellar dust or, mostly in the case of hot Be stars,
due to free-free and bound-free emission from their ionised gaseous
disks. Consequently, the IRAS Point Source Catalogue \citep{IRAS}
became a much used source for the search and identification of stars
surrounded by circumstellar material.  The use of an IRAS two-colour
diagram to systematically identify evolved stars in the IRAS PSC, was
developed and undertaken by \citet{Vn} and \citet{Walk}. They found
that the majority of the infrared (IR) excess stars were carbon or
oxygen rich asymptotic giant branch (AGB) stars undergoing mass
loss. Other studies by \citet{Pl} and \citet{Ju} also discovered
giant and first ascent red giant stars with IR excess.

Another manner of finding objects was to use optical catalogues and
cross-correlate those with the IRAS PSC.  Systematic studies such as
those by \citet{Po}, \citet{St} and \citet{Ou} proved very successful
in returning all well-known Vega-type systems and in addition
uncovering large numbers of both young and evolved stars alike. For
example, the presence of IR excess became one of the defining
characteristics of pre-main sequence Herbig Ae/Be stars, and evolved
post-AGB stars where the AGB mass loss phase had ended and the IR
excess traces the cool, detached dust shell (see for example the
reviews by \citealt{Van} on post-AGB stars, \citealt{Zuck} on dusty
disks and \citealt{Waelk} on Herbig Ae/Be stars).

However, the relatively large beam size of IRAS (45\arcsec x 15\arcsec
at 12$\mu$m and larger at longer wavelengths) meant that regions such
as the Galactic Plane could not be studied properly because of
source confusion. In addition, due to its orbit, the IRAS satellite
also did not observe 4\% of the sky. These so-called ``IRAS gaps''
have never been surveyed in the mid-IR.

The Mid-course Space Experiment (MSX) satellite carried out a mid-IR
survey of the galactic plane and other areas of the sky missed by IRAS
using the SPIRIT III instrument (full details \citealt{Pr}).  The
sensitivity of SPIRIT III at 8$\mu$m is comparable to IRAS, but its
beam size is approximately 35 times smaller (at the longest wavelength)
and is therefore much less hampered by source confusion. It discovered
430~000 objects in the Galactic Plane (defined as $\mid b\mid<$6\degr
), which is four times as many as IRAS detected in this region of the
sky.  Moreover, it surveyed the IRAS gaps for the first time in the
mid-IR.

The MSX survey thus provides an excellent opportunity to
systematically search the Galactic Plane and the IRAS gaps for IR
excess stars.  The purpose of the present paper is to find optically
bright objects within the MSX Point Source Catalogues that were
previously unknown as having IR excess. The resulting list can then be
used for (optical) follow up studies. The methodology we adopt is
similar to that of \citet{Ou}. They cross-correlated the optical SAO
catalogue with the IRAS PSC, to identify objects with IR excess. The
far-IR colours immediately revealed non-photospheric emission if the
temperatures are less than about 200K while those objects with higher
colour temperatures are identified by assessing the IR fluxes compared
to that predicted from the photospheric optical emission. After a
further selection on spectral type (BAFG - typical for most evolved
and young stars) this search revealed a master sample of 462
objects. As MSX did not observe beyond 20$\mu$m, it is not readily
possible to find excess objects based on MSX data alone, and we will
rely on computing the IR excesses using additional information to
identify the stars of interest. Rather than use the SAO catalogue, we
will use the much larger optical Tycho-2 catalogue \citep{Ho} and its
accompanying spectral type catalogue \citep{Wr} which contains more
than 350 000 spectral types as a starting point. To further
characterise the infrared emission of any objects found we will also
use data from the 2MASS All Sky Catalogue of Point Sources (hereafter
2MASS PSC) \citep{Cu}.

This paper is organised as follows. Section \ref{xcorr} describes the
initial input catalogues and the cross correlation of the Galactic
Plane sample in detail. In Section \ref{results} we discuss the
properties of the resulting sample and develop a new method of
identifying IR excess stars. We end with a discussion of the resulting
final sample of excess stars and outline the way it is presented. We
also discuss in Appendix \ref{app} the application of the excess
identification procedure to the MSX IRAS gap survey and other regions
surveyed by the MSX mission.

\section{Input data and cross correlation}
\label{xcorr}

In order to define an IR excess for the sample stars, we use the MSX
mid-IR catalogue, the optical Tycho-2 catalogue and the near-IR 2MASS
PSC. As will be shown later, once the ``optical'' stars have been
identified, the use of their near-IR magnitudes has marked advantages
over using the optical data alone. Below we will describe the input
catalogues and Galactic Plane cross-correlation in detail. The other
MSX catalogue cross-correlations will be discussed in the appendix.

\subsection{Data Sources}
\label{data}

The MSX SPIRIT III instrument observed in six photometric bands
between 4 and 21.34$\mu$m, and was most sensitive at 8.3$\mu$m where
its sensitivity was comparable to the IRAS 12$\mu$m band. The MSX
mission surveyed the galactic plane, IRAS gaps, Magellanic clouds as
well as some other regions of high stellar density. A major data
product of the MSX mission was the MSX PSC v2.3 (hereafter MSX PSC)
\citep{MSXv2.3} which has a positional accuracy, $\sigma=2\arcsec$,
and a limiting flux at 8$\mu$m of 0.1 Jy.

As we wish to identify a large sample of new IR excess stars, we
require a sufficiently large and accurate optical star catalogue. For
this reason we used the Tycho-2 star catalogue compiled from data
collected by the Tycho star mapper on-board the Hipparcos astrometric
satellite.  The positional accuracy of Tycho-2 is excellent and more
than required for our cross-correlation purposes. The completeness
limit (99\%) of the catalogue is given as 11 in the $V_T$ band, and
its limiting magnitude is about 14 in the $V_T$ band and 15 in the
$B_T$ band. Tycho-2 observed objects many times, resulting in a
photometric precision of typically 0.05 mag. Obviously, this precision
becomes worse at the fainter end, for magnitudes around 10-11 the
photometric error is 0.1 mag rising to 0.3 mag close to the detection
limits.

The Tycho-2 catalogue does not contain any stars brighter than
$B_T<2.1$ or $V_T<1.9$ due to the nature of the data reduction
technique. We used the brighter stars from the first supplement to
the Tycho 2 stars catalogue.

Near-IR photometry is taken from the 2MASS PSC which contains nearly
half a billion sources. The survey is complete to magnitude
$K\sim$15. The data has relatively bright saturation limits
($K\sim$3.5), though for $K<8$ the 2MASS catalogue lists photometry
estimated from radial profile fitting to the wings of the saturated
sources. The photometric accuracy, especially at the fainter end, is
better than that of Tycho-2 and for bright saturated sources
comparison with previous observations suggest the error is typically
$<10$\%.

Finally, to learn more about the sample, spectral type information is
very important.  \citet{Wr} compiled all known spectral
classifications for Tycho-2 stars and their catalogue contains
spectral types for 351~863 stars of which 61~472 are in the galactic
plane (defined here as $\mid b\mid<$6\degr ). As might be
expected the objects with spectral types are somewhat brighter on
average than the full sample.

\subsection{Positional Correlation}
\label{correlation}

We have positionally correlated the Tycho 2 star catalogue and the MSX
v2.3 PSC. For a positional association between a Tycho 2 star and a
MSX source, the separation of the two objects is required to be less
than 6\arcsec, corresponding to a typical 3$\sigma$ accuracy of the
MSX position.  In the case of two stars within 6\arcsec \ of an MSX
source we have taken the closest.

The deviations of the MSX positions compared to the Tycho-2 positions
in the right ascension and declination directions are centred at zero
with $\sigma$=2\arcsec \ indicating that, in a statistical sense, the
Tycho-2 and MSX associations are real. The shape of the distribution
compared well with the similar in-scan and cross-scan separation
distributions shown by \citet{MSXv2.3}. To test the 6\arcsec \ cutoff
the correlation was extended out to a radius of 15\arcsec. We found
that the majority (90\%) of the associations had separations smaller
than 6\arcsec \ and the distribution showed that at larger radii the
number of associations reached some random background level. We
therefore concluded that 6\arcsec \ was the optimum value for the
cutoff for positional association.

The MSX-Tycho 2 sources have also been further correlated with the
2MASS PSC, using a 6\arcsec \ search radius around the MSX position.
In the case of multiple 2MASS sources within the 6\arcsec \ search
radius, we have taken the closest match and flagged the MSX source.

To ensure that all MSX-2MASS counterparts are also Tycho 2-MSX-2MASS
associations, we require the 2MASS source to be additionally within
0.50\arcsec \ of the Tycho 2 catalogue position. Failing to do the
latter returns many sources that are either unrelated to the MSX or
the Tycho-2 source and would severely contaminate the final sample.

We made use of the Infrared Science Archive
(IRSA)\footnote{http://irsa.ipac.caltech.edu} to perform the 2MASS
cross-correlations.
 
\subsection{Photometric Constraints}
\label{Photo}

At this stage we made certain requirements of the photometry provided
by the different catalogues for further inclusion in the sample so
presence of the IR excess is credible/significant. For the Tycho 2
star catalogue we required that a star be detected at both the $B_T$
and $V_T$ bands (corresponding to a magnitude less than 15 and 14 in
$B_T$ and $V_T$ respectively according to the Tycho 2 explanatory
supplement \citealt{Ho}). For the MSX data we required a detection in
at least one band with flux quality greater than 1 (signal-to-noise
ratio greater than 5). The only constraint we made on the 2MASS
photometry is that it must be better than an upper limit (i.e not
having a U or X flag) at all bands.

\begin{figure}
\includegraphics[scale=0.75]{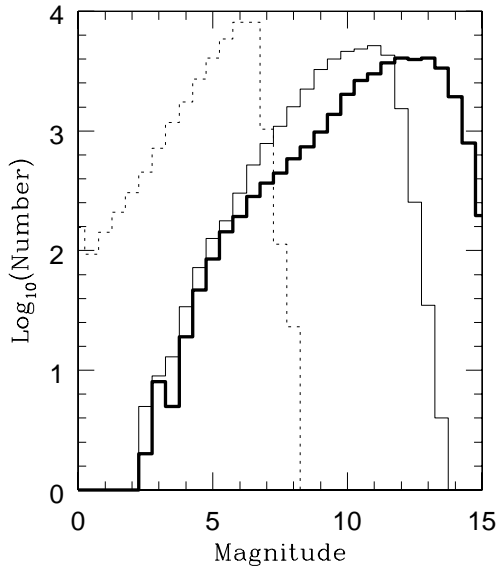}
 \caption{The magnitude distribution of the galactic plane sample in
  the optical and mid-IR. The optical B$_T$ and V$_T$ bands are
  represented by the thick and faint solid lines respectively. The
  MSX 8$\mu$m distribution is indicated by the broken line.}
\label{completeness}
\end{figure}

\subsection{Galactic Plane}
\label{GP}

The MSX PSC (ver 2.3) Galactic Plane $\mid b \mid$ $\leq6$\degr survey
(including the plane regions of the IRAS gaps) contains 431~711
sources. After cross correlation with the Tycho 2 optical star
catalogue, we are left with 35~044 ($\approx 8\%)$ associations which
meet our photometric constraints.  The bulk ($\approx 75\%$) of these
galactic plane sources were only detected at the most sensitive MSX
band A (8.28$\mu$m). The further correlation of the MSX-Tycho 2
sources with the 2MASS PSC search found 33~495 (95\%) counterparts. A
small fraction of these (3\%) had multiple 2MASS sources within the
MSX search area, in these cases we cannot rule out the possibility
that the MSX source is actually a nearby optically invisible 2MASS
source rather than the Tycho 2 star. A clear example of this is the B
type star HD~93942 (the object with $K$$-$$[8]$ $\approx$ 8 in Figure
\ref{nearcolour}) which is in close proximity to a very red (optically
invisible) carbon star. In the near-IR there are two 2MASS sources
within 6\arcsec of the MSX source position. Although the MSX flux is
most likely from the carbon star it is difficult to determine which of
the 2MASS sources is the correct association. As we cannot objectively
remove these sources we flag them.

Using the magnitude distribution of the sample (see Figure
\ref{completeness}) we estimate the sample is complete to magnitude 8
in $V_T$, 11 in $B_T$ and 5 in MSX Band A (or [8]
hereafter\footnote{The MSX fluxes are converted to magnitudes using
zeropoint fluxes from \citet{MSXv2.3} which is 58.49 Jy at
8$\mu$m}). This makes it the most complete IR sample of optical stars
in the galactic plane to date. Of the 35~044 Tycho 2-MSX sources
described above, about a third, 12~783, are listed in the Tycho 2
spectral type catalogue. A smaller fraction of these 7443 have two
dimensional spectral types.

\subsection{Comparison with IRAS}

To ascertain how many of our sources have previously been detected
with IRAS and hence estimate how many objects in our final sample are
new identifications, a positional correlation of the MSX-Tycho 2
sources with the IRAS PSC \citep{IRAS} has also been undertaken.  We
used the VizieR\footnote{http://vizier.u-strasbg.fr} catalogue access
tool to search the IRAS PSC around the MSX positions of our MSX-Tycho
2 associated stars. For the correlation we use a circular search
radius of 45\arcsec \, typically corresponding to three times the
major axis of the elliptically shaped IRAS 1$\sigma$ positional
uncertainties \citep{IRAS}.

Of the 35~044 MSX-Tycho 2 sources, 9473 were found to be within
45\arcsec \ of an IRAS source. We therefore conclude that the majority
of any IR excess stars in this paper are new identifications and
previously unstudied.

\section{Results for the Galactic Plane}
\label{results}

In the following we will first discuss the properties of the sample
using colour-colour diagrams, and then continue with the
identification of excess stars. In order to do this, we will first
derive a relationship between the near-IR colours and the MSX 8$\mu$m
magnitude for normal stars. We will concentrate on the largest MSX
catalogue, that of the Galactic Plane and briefly discuss the
selection from the other samples later.

\subsection{General Properties}
\label{colour}

\begin{figure*}
\includegraphics[scale=0.75]{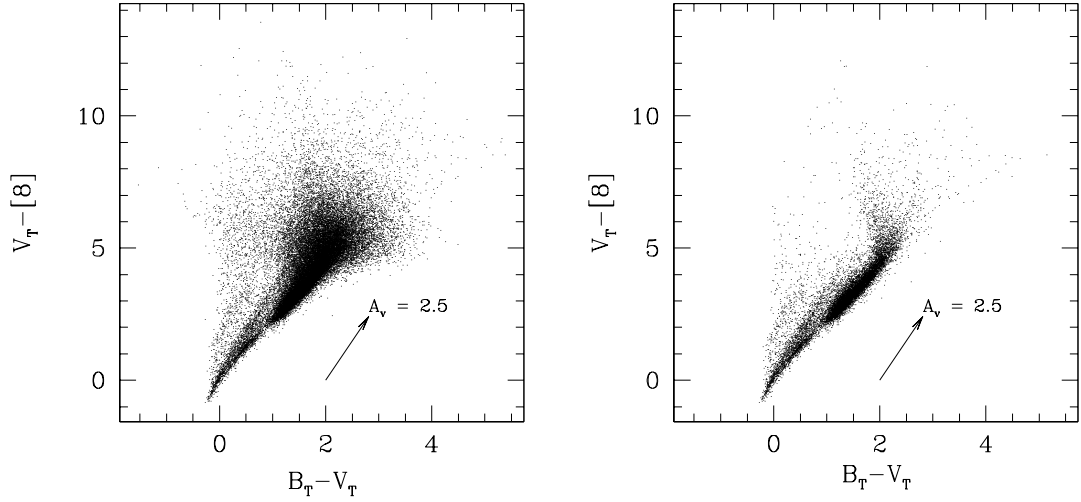}
 \caption{Optical-mid IR colour diagram of the sources. The horizontal
axis denotes the Tycho $B_T-V_T$ colours, while [8] represent the
MSX 8$\mu$m magnitude (see text). The left hand panel shows all 35~030
MSX sources with a Tycho counterpart. The right hand panel shows the
12~778 objects with known spectral types. Note that this (brighter)
sample shows a smaller spread around the main band. For comparison a
reddening vector is shown as well.}
\label{optcolour}
\end{figure*}

Figure \ref{optcolour} shows a ($B_T$$-$$V_T$,$V_T$$-$$[8]$) colour-colour
diagram \footnote{On each diagram the magnitude and direction of a
typical interstellar extinction vector are indicated calculated using
the following assumptions.For the mid-IR wavelengths ($\lambda >
5\micron$) we have used the MSX filter averaged astronomical silicate
data \citet{DraineandLee} as derived by \citet{Lum}. For the near-IR
wavelengths ($\lambda < 5\micron$) we adopt an extinction law that
varies as $\lambda^{-1.75}$. For the optical bands we have used the
standard optical $A_V=3.1(E_{B-V})$ extinction relation.} which is the
MSX analogue to the well studied diagnostic IRAS ({\it
B$-$V},{\it V}$-$[12]) diagram (e.g. \citealt{Wa} herafter WCA).

The left hand panel, containing all objects, shows a main band with
increasing spread around it towards redder colours, the objects with
known spectral types are plotted in the right hand panel. Not
surprisingly this subsample is on average brighter, and a large number
of objects has dropped out, allowing us to recognise a well-defined
band in the plot. The larger spread evident in the full sample is thus
mostly due to larger photometric errors.

This diagram is a good diagnostic to identify IR excess stars, as
originally described by WCA. Normal stars follow a well-defined
sequence, while objects with excess 8$\mu$m emission are readily
identified by their deviating {$V_T$$-$$[8]$} colours. This is also
observed here; the "main band" of stars follows a more or less
well-defined relation and is accompanied by objects located above this
relation. These are the IR excess stars, the majority of stars in the
Tycho 2-MSX sample are not IR excess stars but normal stars.

\begin{figure*}
 \includegraphics[scale=0.99]{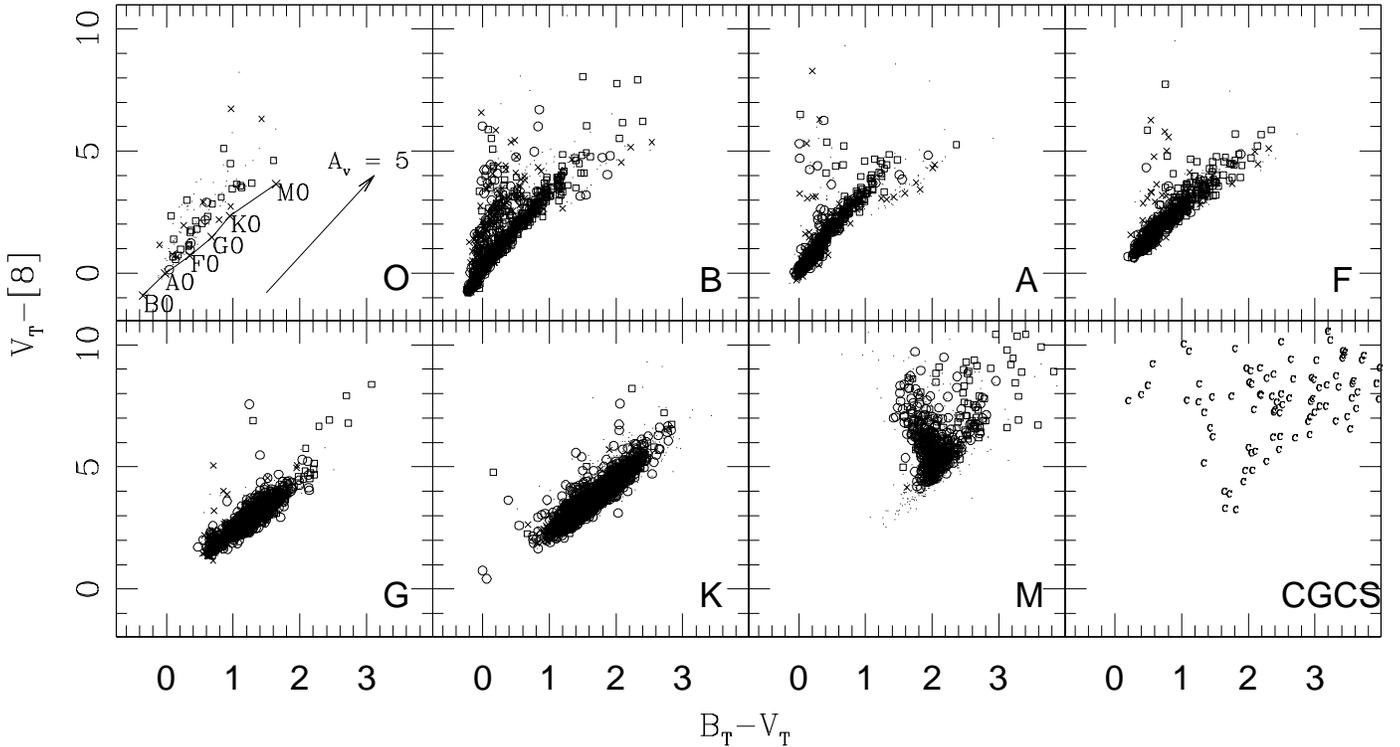}
 \caption{Optical and Mid IR colour diagram showing the different
 regions occupied by different spectral type stars. The spectral type
 is shown in the lower right hand corner of the respective panel. The
 samples consist of 125 O-type stars, 1003 B stars, 718 A stars, 943 F
 stars, 2288 G stars, 6495 K stars and 1099 M stars.  Stars with
 General Catalogue of Galactic Carbon Stars identifications are also
 shown.  The luminosity class of the stars is represented by the
 following symbols $\Box$ for I,II $\circ$ III,IV and $\times$ for V,
 unclassified stars are indicated by a dot. The intrinsic colours of
 dwarf stars (see text for details) together with the extinction
 vector to indicate the degree of reddening of the sample are also
 shown in the upper left hand panel.}
 \label{optspectype}
 \end{figure*}

\subsection{Excess as function of spectral type}

To further investigate the properties of the sample, we plot the
colour-colour diagrams for each spectral type in Figure
\ref{optspectype}. The sample is broken down into the spectral types
O-M while a set of Carbon stars, taken from the Catalogue of Galactic
Carbon Stars (CGCS \citealt{cgcs}), is also shown. Where known, dwarfs
(luminosity class V), giants (IV-III) and supergiants (I-II) are
indicated by different plot symbols.  As a guide for the intrinsic
colours of main sequence stars in Figure \ref{optspectype} we show the
median $V_T$$-$$[8]$ colours of B0 to M0 dwarf stars taken from
\citet{cohen00} plotted against the intrinsic $B$$-$$V$ colours of these
stars given by \citet{kaler} which are corrected for Tycho photometry
using the transformations given in \citet{esa}.

The earlier type stars are dominated by main sequence stars, while
giants constitute the majority for the later type objects as the
dwarfs are too faint to be detected in both optical and IR catalogues.
The transition between dwarfs and giants is visible as the relative
paucity of sources at $ B_T$$-$$V_T$ $\approx$ 0.75.

A separate bifurcation is also visible at $ B_T$$-$$V_T$ $\approx$ 1
in Fig.\ref{optcolour}. The upper sequence consists of reddened
supergiants while the lower sequence consists of giant (and some
dwarfs). This separation of luminosity classes (beginning at early G
type) was first noted by \citet{cohen} in the IRAS $V$$-$$[12]$
diagram. A property apparent for all spectral types is that the
non-excess stars span a much wider $B_T$$-$$V_T$ range than the
intrinsic colours for the respective types signal. This is due to the
presence of a large interstellar reddening at low galactic latitudes,
and is illustrated by the reddening vector indicated in the diagram.
The main bands defined by non-excess stars per spectral type are
parallel to the extinction, and correspond to values up to $A_V \sim
5$. Often the reddest objects are supergiants, which is consistent
with reddening as well; the objects with the largest extinction are
presumably at the largest distances and therefore have to be
intrinsically brighter.

For the B type stars, and also, but less obvious, for the O stars due
to the lower number of stars involved, the colour diagram is dominated
by two bands which are both populated by Main Sequence stars. The
lower band contains normal, non-excess objects, and the upper band is
populated by a large number of excess stars, many of which are Be
stars. The excess emission is explained by free-free emission from the
ionised gas in the circumstellar disk \citep{WaBe}. The relative
number of A type excess stars is much smaller, and seems to consist of
dwarfs and supergiants in roughly equal numbers. There would seem to
be a special class of A type stars around $B_T$$-$$V_T$ = 1.5, where
these seem to branch off at a different slope than the other A type
stars. It appears however that these objects have colours similar to K
type stars and the most likely explanation for this is that their
spectral types are misclassified or that they are binaries with a K
type counterpart.

The F, G and K type stars have fewer excess objects. The M star sample
shows a distinct upturn. As already noted by WCA, for these cool
objects, the optical colours probe the Wien part of the spectral
energy distribution and show hardly any dependency on temperature
anymore. The 8$\mu$m band is sensitive to spectral features, such as
molecular bands and therefore shows a strong gradient. This particular
colour diagram is thus not a useful diagnostic to identify M stars
with IR excess.

Perhaps unsurprisingly, the carbon stars follow a similar band as the
M stars. However the relatively large photometric uncertainties of
these faint stars means they sometimes appear at much bluer colours
than would normally be expected.

It is important to highlight a significant difference with previous
studies here. WCA in-particular, but also \citet{Ou} used optical
catalogues with brighter cut-offs (the Bright Star Catalogue
\citealt{Hof} and the Smithsonian Astrophysical Observatory star
catalogue \citealt{sao} with limits {\it V} = 7 and 9-10 respectively)
than used here. The deeper Tycho-2 sample inevitably contains a large
fraction of (heavily) reddened stars, which is aggravated by the fact
that we observe in the direction of the Galactic Plane.  It can be
seen in the upper left panel of Figure \ref{optspectype}, that the
extinction vector is non-parallel to the ``normal star'' relation,
this gives heavily reddened normal stars $V_T$$-$$[8]$ colours similar
to excess stars.  Moreover, the fainter, numerous, Tycho-2 stars have
comparatively large photometric errors (as indicated by the large
spread in Figure \ref{optcolour}), making it harder to recognise
excess emission from the optical-IR colour-colour diagram. It is
therefore not trivial to properly identify stars with excess from the
Tycho 2-MSX data alone. Therefore, to make headway, we need to go to
wavelengths where the extinction will be less, the near-IR.

\subsection{Near-infrared}

\begin{figure*}
\includegraphics[scale=0.95]{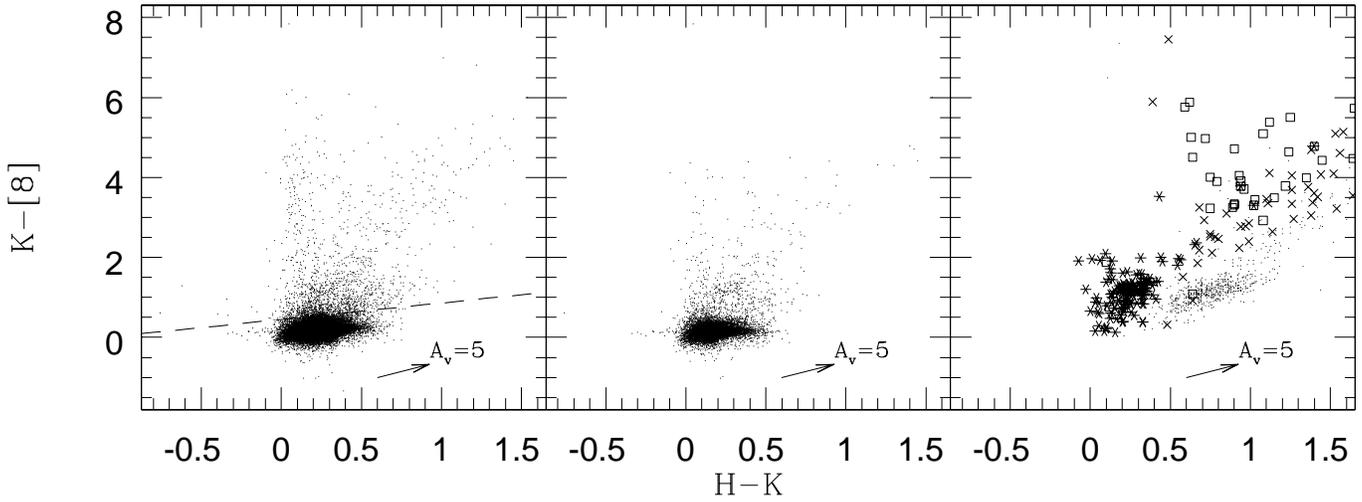}
 \caption{Near-Mid IR (left) colour diagram of MSX-Tycho2-2MASS
sources (33~485), with spectral type (centre) (12~452) and (right) the
colours of known sources (not necessarily in our sample). The excess
cutoff line is shown in the left panel (see text for details). The
symbols in the right hand panel indicate the type of objects ($\ast$
Be type star from our sample, $\times$ a OH/IR star,$\Box$ a Herbig
Ae/Be star and a dot indicates an optical Carbon Star).}
 \label{nearcolour}
 \end{figure*}

To reduce the impact of extinction on the IR excess star selection, we
show the near- and mid-IR colour equivalent to Fig.\ref{optcolour} in
Fig. \ref{nearcolour}. The $K$$-$$[8]$ range spanned by the normal
stars is much smaller than the $V_T$$-$$[8]$ range because we now probe the
Rayleigh-Jeans tail of the spectral energy distribution for all stars.
The $H$$-$$K$ colour range is small for the same reason, while the
reddening is less severe (typically $E(H$$-$$K) = 0.22
E(B$$-$$V)$). Most of the objects beyond $H$$-$$K$$>$0.5 suffer from
reddening, also objects with excess emission at {\it K} give rise to
red $H$$-$$K$ colours.

To understand and describe the features of the colour diagram, we have
also plotted the colours of known objects (Fig.\ref{nearcolour}) which
have been observed by MSX (but are not necessarily part of the Tycho 2
sample).  The objects plotted are from the following catalogues:
Optical selected carbon stars from \citet{cgcs}, OH/IR stars (AGB
stars with large mass loss rates stars) from \citet{cheng}, Herbig
Ae/Be stars from \citet{the} and the Be type stars extracted from our
own Tycho 2 sample (which were found to agree with the mid- near- IR
colours of all known Be stars as compiled by \citealt{Zhang}).

\begin{figure*}
 \includegraphics[scale=0.99]{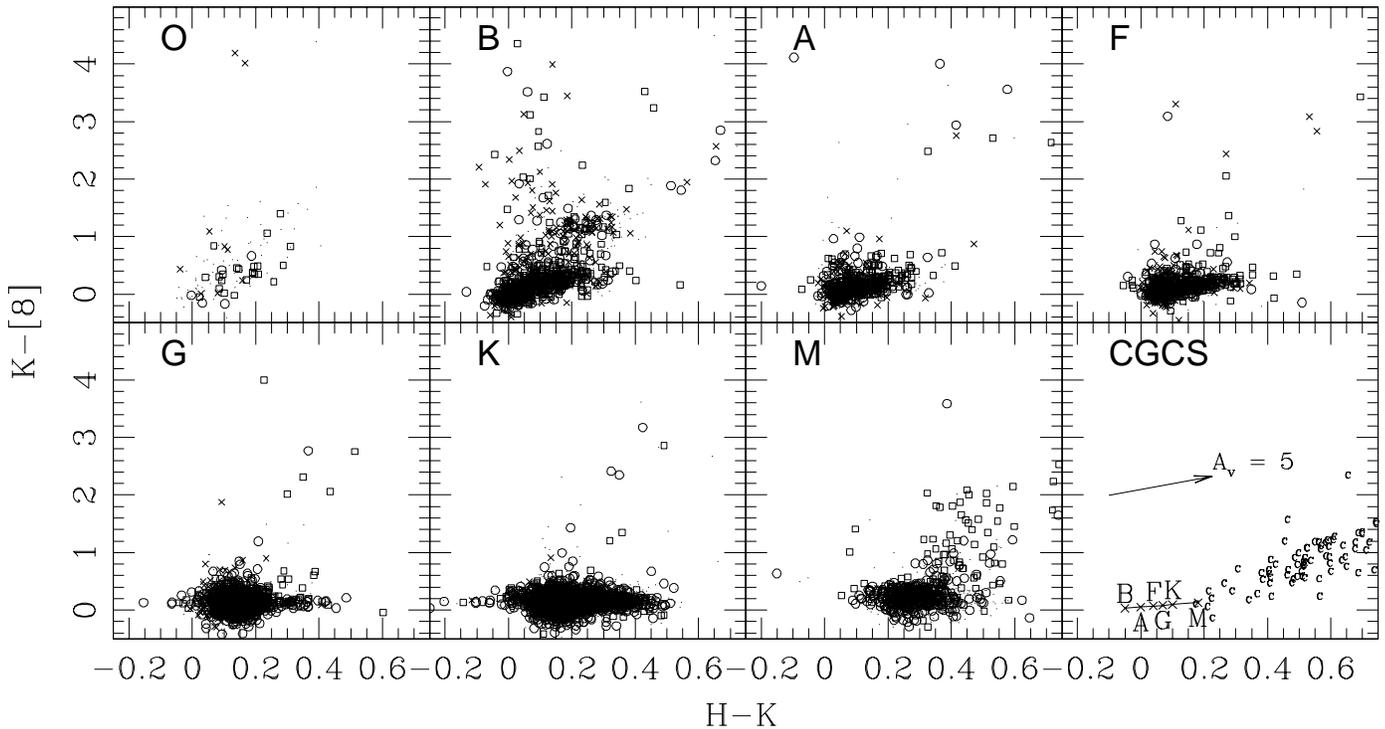}
 \caption{Near-Mid IR colour diagram showing the different regions
 occupied by different spectral type stars. The spectral type is shown
 in the upper left hand corner. The samples consist of 120 O-type
 stars, 937 B stars, 672 A stars, 885 F stars, 2219 G stars, 6406 K
 stars and 1086 M stars. Stars with CGCS identifications are also
 shown.The luminosity classification of the stars is represented by
 the following symbols $\Box$ for I,II $\circ$ III,IV and $\times$ for
 V, unclassified stars are indicated by a dot. The intrinsic colours
 of dwarf main sequence stars (see text for details) together with the
 extinction vector to indicate degree of reddening of the sample are
 are also shown in the lower right panel.}
 \label{nearspectype}
 \end{figure*}

The distribution of B stars with excess in Fig. \ref{nearspectype}
shows a clearly defined band of excess stars with $K$$-$$[8]<2$ and a
more diffuse distribution of stars with greater excess emission, which
is further complemented by a large number of excess stars without
spectral classification (see Fig. \ref{nearcolour}).

It is immediately apparent from Fig.\ref{nearcolour} that only the B
stars with $K$$-$$[8]<2$ can be explained by the properties of the
known Be stars. The remaining objects with greater excess emission are
therefore very interesting; we discuss in section \ref{discuss} the
possible nature of these objects.

The two branches beginning at $H$$-$$K=0.50$, with gradients roughly
equal to the reddening vector , are clearly associated with the
carbon and OH/IR stars. The OH/IR stars also have a slightly greater
(1 mag) $K$$-$$[8]$ colour, presumably due to their higher mass loss
rates over these optically selected carbon stars. The Herbig Ae/Be
stars all show excess emission and are heavily reddened, due to their
surrounding circumstellar material. 

Spot checks on stars with $H$$-$$K<0$ revealed the majority have
saturated {\it K} band magnitudes, and therefore their colours are just
affected by the resulting errorbar. The objects with $K$$-$$[8]<-0.50$
are in general variable stars and binaries.

Although we plot the same number of objects in Figs.\ref{optcolour}
and \ref{nearcolour}, the excess stars are more easily distinguishable
than in the optical diagram. In addition to the reduced extinction,
this is also the case because of the smaller photometric errorbars
involved, validating the use of near-IR colours.

The distribution over spectral type is shown in Figure
\ref{nearspectype}. All features described for the optical
(Fig.\ref{optspectype}) are present, but much more prominent.  It is
clear that the near-mid IR two colour diagram is a much better tool
for the selection of IR excess stars, due to the removal of the strong
extinction effect and the better photometric accuracy of 2MASS for the
optically fainter objects. We therefore proceed with this colour
diagram to identify IR excess stars.

\subsection{The $H$$-$$K$, $K$$-$$[8]$ relation for normal stars}
\label{normal}

To identify stars with excess 8$\mu$m emission, we first need to know
the photospheric contribution for all stars at this wavelength.  To
derive this we adopt the iterative procedure used by WCA for
IRAS sources. This involves dividing the sample into $H$$-$$K$ colour
bins (0.01 mag width in our case) and calculating the mean $(K$$-$$
[8])_{av}$ and standard deviation $\sigma$ of the bin. By eliminating
the outliers (defined as deviating by more than 2$\sigma$ from the
mean), this process was repeated until the ($K$$-$$[8]$)$_{av}$ no longer
changed significantly with further iterations.

The procedure was only run over the intrinsic main sequence near-IR
colours given by \citet{Ko} ($H$$-$$K = -0.05..0.45$).  This step
essentially removes the heavily extincted sources, whose infrared
colours are heavily distorted by reddening.  The final
$(K$$-$$[8])$$_{av}$ values could then be well fitted by the following
straight line.

\begin{equation}
(K-[8])_{photo} = 0.41(H-K)_*+0.05
\label{gpline}
\end{equation}

This photospheric relation was extrapolated to cover the entire range
of ($H$$-$$K$) colours spanned by the sample stars. Using this
relation and the intrinsic main sequence near IR colours quoted by
\citet{Ko} we derive a relationship between the spectral type of a
star and its $K$$-$$[8]$ colour (as shown in the lower right panel of
Figure.\ref{nearspectype}).

\section{Tycho-2 stars with infrared excess}

In this section we select the excess stars. We first discuss the more
numerous 8$\mu$m excess stars, and then discuss excess at the other
MSX wavelengths. 

\subsection{Stars with warm dust : Excess 8 $\mu$m  emission objects}

To calculate the IR excess emission of the stars in our sample we
follow \citet{Ou} who used the relationship between the photospheric
colours in the optical and IRAS 12 $\mu$m for normal stars.  With the
expression of the intrinsic $K$$-$$[8]$ colour for normal stars already in
hand, it is trivial to compute the excess 8$\mu$m emission. Expressed
in magnitudes this is done by $E(K$$-$$[8])=(K$$-$$[8]) - (K$$-$$[8])_{\rm
photo}$.

\begin{figure*}
\includegraphics[scale=0.75]{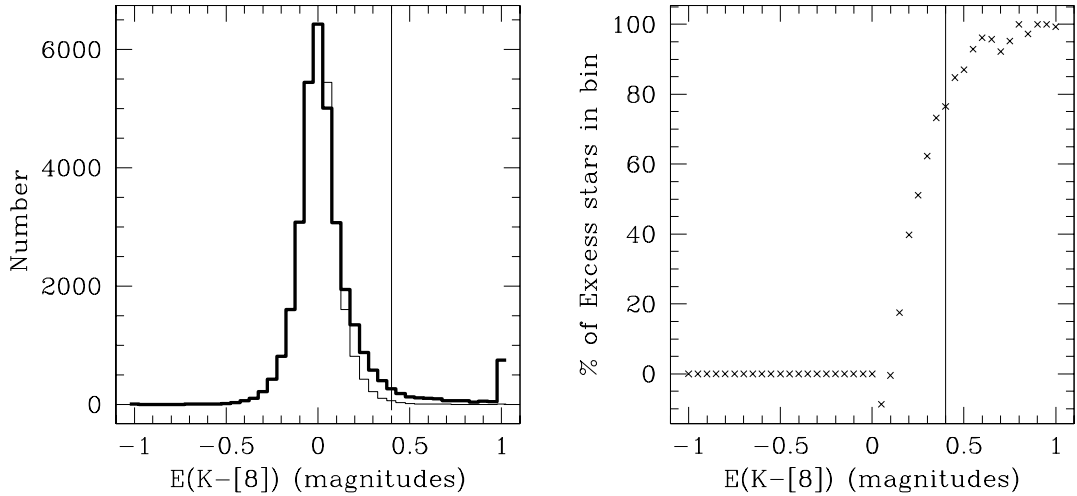}
 \caption{Excess $K-[8]$ ($E(K-[8])$ distribution of sample, the
 negative excess distribution is mirrored through $E(K-[8])=0$ to
 illustrate the asymmetry of the distribution. The right hand panel
 shows the percentage of excess stars as a function of $E(K-[8])$
 emission, the excess definition threshold is also shown at 0.40
 magnitudes.}
 \label{excess}
 \end{figure*}

The main issue is to decide on where to choose the cutoff for
inclusion into the final sample of infrared excess stars. In
Fig.~\ref{excess} we show a histogram of the excesses in magnitudes.
The distribution of $E(K$$-$$[8])$ is asymmetric about zero.  The
negative excesses show a Gaussian shape.  The positive side also
displays an underlying Gaussian distribution, with a large
non-gaussian tail that contains the excess stars. The gaussian shape
on the negative side is explained simply by the width of the observed
sample in the $H$$-$$K$ $K$$-$$[8]$ colour diagram. The width
($FWHM=0.20$) in this case is dominated by the photometric error in
the MSX 8$\mu$m flux (5-20\%) rather than the intrinsic scatter of normal
non-excess stars.

We can now derive the fraction of objects with excess 8$\mu$m
emission.  By using the negative excess distribution (assumed to be
symmetric around zero) as an estimate of the non-excess distribution,
we can calculate the percentage of excess stars as a function of
$E(K$$-$$[8])$. This is shown in Figure \ref{excess}, the percentage of
excess stars is zero for the negative side (by definition) and rises
relatively quickly to 100\% at $E(K$$-$$[8])=0.80$. We choose to have at
least an 80\% probability of IR excess in our sample. This
corresponds to a $E(K$$-$$[8])$ cutoff of 0.40 magnitudes.

Using this criteria we identify 1830 $8\mu$m excess stars in our
sample, around half of which (965) are not within 45\arcsec of an IRAS
Point Source and are therefore new identifications. We also identify
one excess star which was too bright to be included in the Tycho 2
catalogue (and hence was picked up in the Tycho 2 supplement), the
well known Be star $\gamma$ Cas.

\subsection{Stars with hot dust : $K$ band excess emission objects}

It should be noted that the above procedure selects objects that have
excess 8$\mu$m emission {\it relative to} the {\it K} band. An object
surrounded by only hot dust will not be selected as its 8$\mu$m
emission has a similar $K$$-$$[8]$ colour as a cool C or M type
star. Indeed, the intrinsic $K$$-$$[8]$ colour for a 1000-2000 K Black
Body is about $\sim$ 0.5, and it may well be that such objects will
not be selected as their observed $K$$-$$[8]$ colour seems to all
intents and purposes "normal".  We therefore need to identify {\it K}
band excess stars separately.

\begin{figure*}
\includegraphics[scale=0.75]{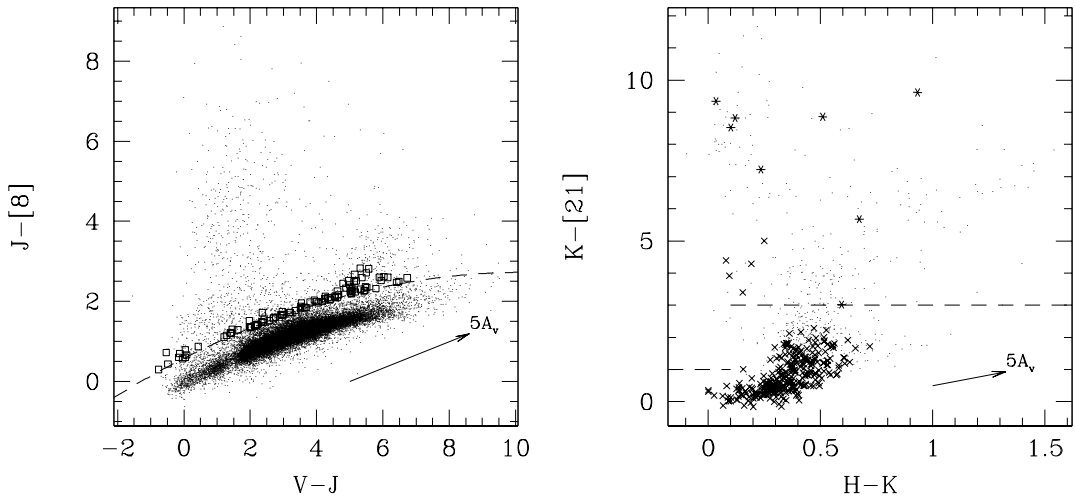}
\caption{Optical Mid infrared and Near-Mid infrared colour diagrams
showing the J-[8] and K-[21] colours for hot and cold excess emission
respectively. In the left panel, a $\Box$ indicates a star with excess
J-[8] emission but normal K-[8] colours, the rest of the sample is
indicated by a dot. The right panel shows the K-[21] colours for cool
excess identification. Stars detected at 21$\mu$m but not at 8$\mu$m
are indicated by a $\ast$, stars detected at both 8$\mu$m and 21$\mu$m
are indicated by a dot if they show excess at 8$\mu$m or a $\times$ if
they do not}
\label{hotcoldexcess}
\end{figure*}

We investigated several ways to identify such objects and found the
best method is to use $V_T$$-$$J, J$$-$$[8]$ colours as shown in
Figure.\ref{hotcoldexcess}. This avoids the problem of having the
large errorbars we encountered in the optical $B_T$$-$$V_T$ colours
and includes objects with excess even at the {\it H} band. It is
extremely rare for hot dust to be responsible for {\it J} band
excess. Indeed, checks on a random sample of {\it J} excess emission
objects revealed them to be uncatalogued binary objects, where the
{\it J} emission is due to a bright cool secondary.  We identify the
{\it K} band excess stars using a similar iterative procedure as for
the $K$$-$$[8]$ colours and we derive the following relation for the
photospheric contribution.

\begin{equation}
(J-[8])_{photo}=-0.025+0.420(V_T-J)-0.021(V_T-J)^2
\label{hotms}
\end{equation}

The excess $J$$-$$[8]$ colour distribution showed that the probability
of excess reached 80\% at a $E(J$$-$$[8]) = 0.60$ magnitudes and this
is where we placed the excess cutoff. This method of excess
identification returned only 68\% of the $K$$-$$[8]$ excess sources
and a further 95 stars which do not have excess compared to {\it K} at
$8\mu$m or $21\mu$m. Interestingly only 9 were not observed by IRAS,
possibly because the hot excess stars are relatively infrared bright.

\subsection{Stars with cool dust : 21 $\mu$m excess emission objects}

Stars such as post-AGB stars surrounded by a cool detached dust shell
may not show any distinguishing excess emission at 8$\mu$m. We
therefore wish to use the longer wavelength MSX bands to identify such
sources in our sample.  The previous searches for cold excess objects
(e.g \citealt{Ou}) made use of the IRAS far-IR colour-colour diagram
to identify excess if temperatures are less than 200K (indicating that
infrared emission was not photospheric). A similar approach with MSX
is hampered by the reduced sensitivity of MSX at longer
wavelengths. Indeed, very few sources (2\%) in the sample are detected
(i.e flux quality greater than 1) in every band. As a compromise we
will use the $K$$-$$[21]$ colour (see Figure \ref{hotcoldexcess}) to
identify cool excess objects. This has the further advantage that
stars detected exclusively at $21\mu$m are not removed from the sample
as they would be if we relied on a multi MSX colour diagram.

Figure \ref{hotcoldexcess} shows the $K$$-$$[21]$ colours, of all
sources with a flux quality greater than 1 at 21$\mu$m, plotted
against {\it H$-$K}. Objects that are found to display 8$\mu$m excess
emission are indicated by a dot, while those not detected at 8$\mu$m
or not found to have excess are indicated by larger plot symbols.  It
is immediately apparent that the majority of the cool excess objects
also have excess $8\mu$m emission.

Of the stars without 8$\mu$m excess only a few have $K$$-$$[21]$
colours significantly different from the rest of the sample (the
majority of which are M giants). All sources that are detected at
$21\mu$m but are not detected in the more sensitive $8\mu$m band show
cool excess emission. To determine the excess cutoff we have used the
median $V$$-$$[25]$ values published by \citet{cohen} and converted to
$K$$-$$[25]$ using the appropriate $V$$-$$K$ colours from
\citet{Ko}. As the [21] and [25] bands both probe the molecular
absorption bands of the cool M type giant stars, we may assume that
the $K$$-$$[25]$ and $K$$-$$[21]$ colours are roughly comparable. We
therefore define the excess cutoff to be the 3 sigma deviation from
the intrinsic $K$$-$$[25]$ colours of giants. Hence we select stars
with cool infrared excess if they have $H$$-$$K<0.1$ and $K$$-$$[21]>1$
or $K$$-$$[21]>3$ at $H$$-$$K>0.1$, as shown in Fig
\ref{hotcoldexcess}. Using this criteria we identify 13 cool IR excess
stars which do not have excess 8$\mu$m emission, 5 of which do not
have IRAS counterparts.

\section{Discussion}
\label{discuss}
In this paper we have searched for stars with infrared excess in the
combined Tycho-2, 2MASS and MSX catalogues. This resulted in a grand
total of 1938 excess stars, of which 979 were not detected by IRAS and
are thus newly discovered objects.

\subsection{Distribution over Spectral Type}

\begin{figure*}
\includegraphics[scale=0.75]{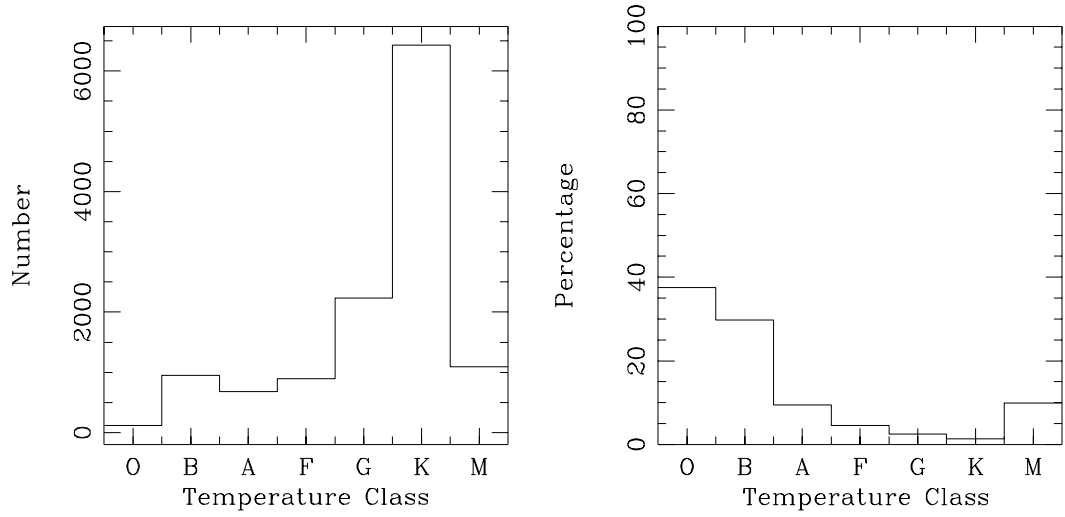}
\caption{Temperature class distribution of entire MSX-Tycho 2 sample
(left) and percentage of stars with excess emission as a function of
spectral type (right).}
\label{spechist}
\end{figure*}

In Figure \ref{spechist} we show the number of Tycho 2-MSX stars as a
function of temperature class. Alongside we also show the percentage
of these stars which were found to have excess emission.

We find that the percentage of excess appears to decrease with
effective temperature and that the hotter stars (O,B and A) and very
cool (M) have a much higher percentage of excess stars than the F,G
and K classes. However it is interesting to note that the numbers of
Tycho 2-MSX stars detected shows the exact opposite behaviour,
indicating that a strong selection effect may be present.

A comparison of the Tycho 2-MSX sample with the entire galactic plane
Tycho 2 spectral type catalogue shows that we detect a higher fraction
of G and K type compared to other spectral types. The most obvious
source of this effect is the sensitivity of the MSX satellite. For
example, the photospheric emission at $8\mu$m for all but the
brightest O,B and A type stars may be below the detection threshold of
MSX; we only detect the excess and bright non-excess stars
(which are relatively few) thus leading to a higher fraction of excess
stars for these spectral types.

\subsection{B type stars with infrared excess}

The high fraction (30\%) of excess in B type stars is surprising, as
the fraction of Be type stars in the Bright Star Catalogue is well
known to be 15\% \citep{coteBe}. We also noted earlier that there
exists a large group of B type (with and without spectral
classification) excess stars with colours that were significantly
different to the known Be stars. The nature of these objects is not
altogether obvious.

Their $H$$-$$K$ colours are quite blue indicating that these are
indeed B type stars. The normality of the $H$$-$$K$ colour indicates
that they do not have a great deal of {\it K} band excess. The absence
of excess in the {\it K} band would seem to lead us away from a
free-free emission explanation for these excesses and indicates that
these stars are unrelated to the Be stars immediately below them in
Figure \ref{nearcolour}. The objects are also not particularly
reddened, ruling out heavily dust embedded objects and implying the
surrounding dust must be relatively optically thin.

We are therefore looking for a optically thin thermally radiating dust
mechanism for these stars. Possibilities include: Hot Post-AGB stars
with a optically thin cool detached dust shell, reflection
nebulosities, giant stars heating the surrounding interstellar medium
dust, weak HII regions and Vega type stars surrounded by a close warm
disk of dust.

To resolve the situation we looked at the spectral energy
distributions of the subsample with IRAS detections. We showed that
the objects with intermediate $K$$-$$[8]$ colours ($2-4$) to be Herbig
stars, Be stars, reflection nebulae and post-AGB stars. The objects
with larger $K$$-$$[8]$ colours were found to be predominately objects
such as post-AGB stars, HII regions and a number of galactic cirrus
contaminated fields (some possibly also heated by giant stars). We did
not find much evidence for Vega type disk systems. To complement this
we also inspected Galactic Legacy Infrared Mid-Plane Survey
Extraordinaire (GLIMPSE; see \citealt{glimpse}) images for the small
number ($\approx 10$) of these stars which were within the survey
region. The GLIMPSE images mostly showed these objects to be stars
within diffuse emission regions and/or associated with star forming
regions. These objects would benefit from ground based optical follow
up observations to better confirm their nature and to further
understand this region of the colour diagram.

\subsection{Presentation of the Data: Data Tables}

We publish in hardcopy (see Table \ref{finaltable}) the stars which
were not detected by IRAS (and therefore should be new
identifications), that have full spectral type information and excess
$K-[8]>0.75$ mag and/or excess emission at other bands. The entire
sample (including the other regions discussed in the Appendix) is
published in the online version of this paper (see Tables 2,3,4 and
5). A machine readable format will also be made available at CDS
Strasbourg \footnotemark[3].

\subsection{Final Remarks}

To identify stars with IR excess emission we have positionally
correlated the MSX PSC with the Tycho 2 star catalogue. We found that
a near-mid IR colour diagram had marked advantages over a optical-mid
IR colour diagram in selecting these objects, specifically in reducing
the strong line of sight extinction experienced in the direction of
the galactic plane. Using the derived colours of normal stars a
selection criteria was developed to identify excess emission sources
from the colour diagram. The criteria produced a sample of 1938 stars
in the galactic plane, just over 50\% of which were determined to be
new identifications of infrared excess (i.e not previously detected by
IRAS).

The majority of the excess stars were found to be hot stars, a high
(30\%) fraction of B type stars with excess was discovered. The
infrared excesses for a number of these B type stars were found to be
much greater than those of the known Be sample. The known objects in
this group were found to be a mixture of Herbig stars, post-AGB stars,
reflection nebulae and stars contaminated by galactic cirrus.

The other regions surveyed by MSX were also searched for IR excess
stars and the same selection criteria were applied (see Sec
\ref{app}). We publish the entire excess sample online (see Table 2)
and the brightest new identifications of excess stars in the galactic
plane, in Table \ref{finaltable}. These lists should provide a good
starting point for future ground based follow up observations.

\subsubsection*{Acknowledgements}

We are grateful to the Royal Astronomical Society who made the pilot
study leading to this work possible by the provision of a Summer
Student grant to AJC and PPARC for the PhD student grant. We would
also like to thank the referee, Michael Egan for many useful comments.
This publication makes use of data products from the Two Micron All
Sky Survey, which is a joint project of the University of
Massachusetts and the Infrared Processing and Analysis
Center/California Institute of Technology, funded by the National
Aeronautics and Space Administration and the National Science
Foundation. This research has made use of the NASA/ IPAC Infrared
Science Archive, which is operated by the Jet Propulsion Laboratory,
California Institute of Technology, under contract with the National
Aeronautics and Space Administration.  This research has made use of
the SIMBAD database, operated at CDS, Strasbourg, France.

\newpage
\subsubsection*{Notes to Table 1}

Table 1 contains: the Tycho identifier, Tycho (J2000) position, HD
name from the HD catalogue identifications for Tycho 2 stars
\citep{hdnames}, spectral type from the Tycho 2 spectral type
catalogue \citep{Wr} , Tycho $B_T$ and $V_T$ band magnitudes, 2MASS
J,H,K near-IR magnitudes, a flag indicating the number of extra 2MASS
sources within 6\arcsec of the MSX position (\# 2MASS), the MSX
8,12,14 and 21 band fluxes (in Jansky), the photometric quality of
each of the MSX bands (upper limit=1 to excellent=4) and finally the
excess colour in magnitudes. The excess colour listed is $K$$-$$[8]$
unless the entry has superscript J or 21, in which case it is excess
$J$$-$$[8]$ or $K$$-$$[21]$ respectively. The typical errors in the
Tycho 2 photometry are 0.05 mag rising to 0.3 mag close to the
detection limits.  The 2MASS photometry is of higher accuracy than
Tycho 2 especially at the fainter end.

\subsubsection*{Notes to the  Online only Tables 2,3,4,5 and 6}

The tables 2,3,4,5 and 6 are only available in the online version of
this paper. They have the same format as Table 1, aside from an extra
column indicating the presence of an IRAS counterpart. The entry
contains an 'I' if an IRAS source was found within 45\arcsec and is
empty if none was found. Additionally Table 6 also includes a column
indicating which star forming region the source is associated with.

\begin{onecolumn}
\begin{landscape}
\pagestyle{empty}
\setlength\tabcolsep{5.0pt}
\begin{longtable}{@{}rrrccrlrrrrrcrrrrrr}
\caption{Tycho 2 stars with infrared excess}\\
\hline
TYC1 & TYC2 & TYC3 & Ra & Dec & HD & SpType &  $B_T$ & $V_T$ & $J$ & $H$ & $K$ & $\#$ & F8 & F12 & F14 & F21 & Flux & Excess \\
 & & & (J2000)&(J2000)  & & & mag. & mag. & mag. & mag. & mag. &2MASS & Jy & Jy & Jy & Jy & Quality & mag. \\ 
\hline
\endfirsthead
\caption{(continued)}\\
\hline
TYC1 & TYC2 & TYC3 & Ra & Dec & HD & SpType &  $B_T$ & $V_T$ & $J$ & $H$ & $K$ & $\#$ & F8 & F12 & F14 & F21 & Flux & Excess \\
& & & (J2000)&(J2000)  & & & mag. & mag. & mag. & mag. & mag. &2MASS & Jy & Jy & Jy & Jy & Quality & mag.  \\ 
\hline
\endhead
\hline
\endfoot
\hline
\endlastfoot
3664  &  192  &  1&  00 11 37.1  &  +58 12 42  &  698  &   B5II:  &  7.26  &  7.11  &  6.49  &  6.43  &   6.33  &  0  &  0.39  &     &     &      &  4 0 0 0  &  0.79  \\  
4034  &  1319  &  1&  01 24 19.4  &  +62 49 36  &     &   B1:PE(V)  &  11.51  &  10.63  &  8.43  &  8.13  &   7.81  &  1  &  0.15  &     &     &      &  3 0 0 0  &  1.17  \\  
4032  &  17  &  1&  01 58 39.1  &  +60 37 43  &     &   B 1 VE  &  10.81  &  10.23  &  8.32  &  8.06  &   7.74  &  0  &  0.16  &     &     &      &  3 0 0 0  &  1.15  \\  
3697  &  1551  &  1&  02 02 36.4  &  +59 41 17  &  12302  &   B1:V:PE  &  8.37  &  8.12  &  7.22  &  7.09  &   6.90  &  0  &  0.33  &     &     &      &  4 0 0 0  &  1.16  \\  
4041  &  1668  &  1&  02 15 13.0  &  +64 01 28  &  13590  &   B2III  &  8.33  &  8.00  &  7.14  &  6.99  &   6.83  &  0  &  0.29  &     &     &      &  3 0 0 0  &  0.93  \\  
3694  &  1707  &  1&  02 22 06.4  &  +57 05 25  &     &   B 2 III-IVE  &  10.05  &  9.83  &  8.48  &  8.10  &   7.55  &  0  &  0.29  &     &     &      &  4 0 0 0  &  1.53  \\  
4049  &  1502  &  2&  03 16 17.4  &  +60 02 07  &  20053  &   B 3 V  &  8.43  &  8.31  &  7.98  &  8.02  &   8.00  &  1  &  0.17  &  2.96  &     &      &  2 1 0 0  &  1.61  \\  
3718  &  756  &  1&  04 04 21.6  &  +53 19 44  &  25348  &   B1VPNNE  &  8.44  &  8.20  &  7.34  &  7.16  &   6.94  &  0  &  0.23  &     &     &      &  4 0 0 0  &  0.77  \\  
2401  &  381  &  1&  05 09 56.4  &  +37 00 16  &  33152  &   B 1 VE  &  8.37  &  8.15  &  7.69  &  7.66  &   7.53  &  0  &  0.25  &     &     &      &  3 0 0 0  &  1.51  \\  
2900  &  503  &  1&  05 10 48.2  &  +41 00 10  &  33232  &   B 2 VNE  &  8.29  &  8.25  &  7.91  &  7.81  &   7.64  &  1  &  0.14  &     &     &      &  2 0 0 0  &  0.97  \\  
2900  &  112  &  1&  05 13 13.3  &  +40 11 36  &  33604  &   B2V:PE  &  7.37  &  7.39  &  7.44  &  7.46  &   7.49  &  0  &  0.18  &     &     &      &  3 0 0 0  &  1.15  \\  
2411  &  16  &  1&  05 28 07.1  &  +34 25 26  &     &   B 0 IV  &  9.53  &  9.40  &  8.38  &  8.36  &   8.32  &  0  &  0.16  &  0.92  &     &   3.44  &  3 1 0 1  &  1.85  \\  
2416  &  541  &  1&  05 40 59.5  &  +35 50 46  &     &   O 9.5V)  &  11.64  &  10.67  &  8.38  &  8.12  &   7.95  &  1  &  1.55  &  4.57  &  7.26  &   19.67  &  4 4 4 4  &  3.89  \\  
1871  &  1417  &  1&  05 53 06.1  &  +26 26 43  &  39340  &   B3V  &  8.18  &  8.11  &  7.59  &  7.58  &   7.26  &  0  &  0.23  &     &     &      &  4 0 0 0  &  1.06  \\  
1871  &  1264  &  1&  05 53 59.8  &  +26 25 21  &  39478  &   B 2 V  &  8.34  &  8.27  &  7.81  &  7.69  &   7.56  &  0  &  0.15  &     &     &      &  3 0 0 0  &  1.01  \\  
1867  &  119  &  1&  05 59 53.6  &  +25 05 19  &  250028  &   B2:V:PNNE  &  9.35  &  9.14  &  8.23  &  8.08  &   8.16  &  1  &  0.19  &     &     &      &  3 0 0 0  &  1.88  \\  
1321  &  1207  &  1&  06 00 11.6  &  +19 11 33  &  250163  &   B1,5:V:PNE  &  10.40  &  9.83  &  8.00  &  7.76  &   7.44  &  0  &  0.18  &     &     &   3.00  &  2 0 0 1  &  0.98  \\  
1877  &  931  &  1&  06 08 35.1  &  +22 37 01  &  41870  &   F8IB-G5IB  &  10.64  &  9.18  &  6.74  &  6.32  &   6.13  &  0  &  0.58  &     &     &      &  4 0 0 0  &  0.98  \\  
1315  &  1999  &  1&  06 23 24.7  &  +15 06 06  &  44637  &   B2V  &  8.27  &  8.05  &  7.33  &  7.25  &   7.06  &  0  &  0.24  &     &     &      &  4 0 0 0  &  0.97  \\  
750  &  899  &  1&  06 41 05.9  &  +09 22 55  &  261941  &   A 5 IV  &  11.33  &  11.32  &  10.55  &  10.15  &   9.57  &  0  &  0.23  &     &  0.89  &      &  3 0 1 0  &  3.27  \\  
4808  &  2967  &  1&  06 47 56.9  &  -05 09 14  &  49370  &   B8/9 V  &  7.57  &  7.62  &  7.70  &  7.75  &   7.71  &  0  &  0.11  &     &     &      &  2 0 0 0  &  0.81  \\  
5381  &  2964  &  1&  07 05 35.2  &  -08 43 43  &  53667  &   B0/1 Ib  &  7.96  &  7.76  &  7.13  &  7.10  &   6.92  &  0  &  0.23  &     &     &      &  4 0 0 0  &  0.79  \\  
5398  &  3009  &  1&  07 11 20.8  &  -10 25 43  &  55135  &   B3 Vnne  &  7.21  &  7.31  &  7.38  &  7.37  &   7.29  &  0  &  0.28  &     &     &      &  4 0 0 0  &  1.42  \\  
5399  &  3900  &  1&  07 22 24.8  &  -10 49 21  &  57775  &   B1/2 (I)ne  &  9.47  &  9.37  &  8.34  &  8.13  &   7.81  &  0  &  0.12  &     &     &      &  2 0 0 0  &  0.93  \\  
5966  &  2402  &  1&  07 27 51.0  &  -16 05 37  &  59094  &   B2 (V)NE  &  8.63  &  8.50  &  7.87  &  7.80  &   7.54  &  0  &  0.16  &     &     &      &  3 0 0 0  &  0.98  \\  
7119  &  178  &  1&  07 49 06.0  &  -31 07 42  &     &   F2EIAB  &  10.40  &  9.52  &  7.51  &  7.10  &   6.82  &  0  &  0.38  &     &     &      &  4 0 0 0  &  1.19  \\  
7683  &  1320  &  1&  08 49 31.7  &  -41 33 49  &  75607  &   B3 V  &  8.85  &  8.77  &  8.40  &  8.35  &   8.44  &  1  &  0.19  &     &     &      &  3 0 0 0  &  2.19  \\  
7689  &  1211  &  1&  09 00 22.3  &  -43 10 26  &  77320  &   B2 VNN(E)  &  5.84  &  6.03  &  6.18  &  6.21  &   6.05  &  0  &  0.34  &     &     &      &  4 0 0 0  &  0.69$^{J}$  \\  
8585  &  3781  &  1&  09 44 07.2  &  -53 44 43  &  84523  &   B3/5 VE  &  8.01  &  7.96  &  7.61  &  7.53  &   7.33  &  0  &  0.20  &     &     &      &  4 0 0 0  &  1.01  \\  
8609  &  2680  &  1&  10 29 47.2  &  -56 36 46  &     &   B 0 VE  &  10.66  &  10.12  &  8.57  &  8.40  &   8.02  &  0  &  0.14  &  0.69  &     &      &  2 1 0 0  &  1.27  \\  
8626  &  2201  &  1&  10 46 12.7  &  -59 19 05  &     &   A6/7 IV/V  &  8.83  &  8.55  &  7.25  &  7.17  &   7.14  &  0  &  0.20  &     &     &      &  4 0 0 0  &  0.90  \\  
8957  &  2047  &  1&  10 47 38.9  &  -60 37 04  &  93683  &   B0/1 (V)NE  &  8.04  &  7.98  &  7.36  &  7.26  &   7.05  &  0  &  0.25  &     &     &   3.35  &  4 0 0 1  &  0.97  \\  
8963  &  1364  &  1&  11 20 27.9  &  -61 52 36  &  98678  &   M5 (III)E  &  11.58  &  9.89  &  5.41  &  4.47  &   3.73  &  0  &  8.60  &  5.59  &  4.17  &   3.12  &  4 4 4 3  &  1.29  \\  
8977  &  3691  &  1&  11 47 35.8  &  -63 22 38  &  309036  &   B 2 II  &  10.92  &  10.77  &  9.97  &  9.93  &   9.81  &  1  &  0.16  &     &     &      &  3 0 0 0  &  3.33  \\  
8985  &  2714  &  1&  11 48 02.2  &  -66 06 53  &  102579  &   K0 V  &  9.65  &  8.65  &  7.08  &  6.70  &   6.57  &  0  &  0.32  &     &     &   3.01  &  4 0 0 1  &  0.80  \\  
8974  &  1391  &  1&  12 03 32.6  &  -61 05 53  &  104722  &   B2 VNE  &  7.53  &  7.57  &  7.10  &  7.07  &   6.94  &  0  &  0.22  &     &     &      &  3 0 0 0  &  0.76  \\  
8645  &  2059  &  1&  12 14 01.8  &  -59 23 48  &  106309  &   B2 III/VNE  &  7.78  &  7.86  &  7.82  &  7.83  &   7.62  &  0  &  0.18  &     &     &      &  2 0 0 0  &  1.17  \\  
8994  &  1920  &  1&  13 10 20.5  &  -62 41 18  &  114214  &   F0 V  &  10.15  &  9.39  &  7.73  &  7.50  &   7.35  &  0  &  0.19  &     &     &      &  3 0 0 0  &  1.00  \\  
8994  &  2766  &  1&  13 14 11.5  &  -63 22 25  &  114800  &   B2 III/VNE  &  8.05  &  7.98  &  7.51  &  7.36  &   7.10  &  0  &  0.25  &     &     &      &  4 0 0 0  &  1.01  \\  
8994  &  3012  &  1&  13 20 35.5  &  -63 24 43  &  115746  &   B2/4 III/V  &  9.62  &  9.38  &  8.08  &  7.92  &   7.67  &  0  &  0.14  &     &     &      &  2 0 0 0  &  0.95  \\  
8990  &  3680  &  1&  13 22 37.9  &  -60 59 18  &  116087  &   B3 V  &  4.34  &  4.49  &  5.24  &  4.90  &   4.89  &  0  &  0.61  &     &     &      &  4 0 0 0  &  0.65$^{J}$  \\  
8995  &  3220  &  1&  13 27 25.1  &  -62 38 56  &  116781  &   O9/B1 (I)E  &  7.74  &  7.67  &  7.22  &  6.99  &   6.72  &  0  &  0.44  &     &     &      &  4 0 0 0  &  1.23  \\  
9003  &  1626  &  1&  13 28 07.3  &  -66 16 47  &  116849  &   B1 VPE  &  9.33  &  9.17  &  8.72  &  8.52  &   8.29  &  0  &  0.13  &     &     &      &  2 0 0 0  &  1.49  \\  
8999  &  53  &  1&  13 29 54.6  &  -65 30 07  &  117111  &   B2 (V)NE  &  7.78  &  7.72  &  7.28  &  7.17  &   7.04  &  0  &  0.24  &     &  0.51  &      &  4 0 1 0  &  0.98  \\  
9016  &  519  &  1&  13 45 18.4  &  -66 45 16  &  119423  &   B3/5 VNE  &  7.54  &  7.59  &  7.59  &  7.62  &   7.59  &  0  &  0.12  &     &     &      &  2 0 0 0  &  0.81  \\  
8676  &  1771  &  1&  13 50 26.1  &  -59 44 52  &  120330  &   B2/3 VNNE  &  7.93  &  7.91  &  7.51  &  7.41  &   7.22  &  0  &  0.25  &     &     &      &  4 0 0 0  &  1.16  \\  
9005  &  873  &  1&  14 15 09.9  &  -61 06 42  &  124340  &   A0 V  &  8.42  &  8.26  &  7.77  &  7.75  &   7.68  &  0  &  0.14  &     &     &      &  2 0 0 0  &  1.02  \\  
9007  &  747  &  1&  14 35 33.3  &  -61 00 29  &  127756  &   B1/2 VNE  &  7.69  &  7.60  &  7.06  &  6.93  &   6.66  &  0  &  0.34  &     &     &      &  4 0 0 0  &  0.92  \\  
8692  &  1383  &  1&  14 42 05.6  &  -58 21 34  &  128937  &   B3/5 (III)  &  7.88  &  7.72  &  7.34  &  7.34  &   7.24  &  0  &  0.35  &     &     &      &  4 0 0 0  &  1.58  \\  
9024  &  980  &  1&  14 59 21.9  &  -62 30 32  &  131925  &   A3 IV  &  8.51  &  8.20  &  7.45  &  7.34  &   7.23  &  0  &  0.19  &     &     &      &  3 0 0 0  &  0.89  \\  
9025  &  1827  &  1&  15 09 03.3  &  -61 53 15  &  133738  &   B1/2 III/VNE  &  6.93  &  6.97  &  6.66  &  6.62  &   6.41  &  0  &  0.47  &     &     &      &  4 0 0 0  &  1.04  \\  
8702  &  636  &  1&  15 10 59.6  &  -57 42 43  &     &   O 9.5IA  &  11.51  &  10.54  &  7.65  &  7.35  &   7.12  &  0  &  0.22  &     &     &      &  4 0 0 0  &  0.90  \\  
8706  &  14  &  1&  15 15 16.2  &  -58 10 22  &  134958  &   B2 (II)NE  &  8.48  &  8.20  &  6.76  &  6.57  &   6.34  &  0  &  0.44  &     &     &      &  4 0 0 0  &  0.88  \\  
8695  &  2105  &  1&  15 25 55.3  &  -53 46 15  &  136968  &   B5 VNE  &  8.26  &  8.25  &  8.34  &  8.28  &   8.14  &  1  &  0.12  &  1.85  &     &      &  2 1 0 0  &  1.33  \\  
8304  &  641  &  1&  15 42 19.8  &  -49 59 52  &  139790  &   B2 (III)N  &  8.62  &  8.57  &  7.92  &  7.76  &   7.55  &  0  &  0.16  &     &     &      &  2 0 0 0  &  1.02  \\  
8321  &  711  &  1&  15 46 47.4  &  -52 08 46  &  140605  &   B5/7 IV  &  7.01  &  7.07  &  6.96  &  6.95  &   6.86  &  0  &  0.34  &     &     &      &  4 0 0 0  &  1.19  \\  
8697  &  718  &  1&  15 55 29.6  &  -52 48 15  &  142170  &   B8/9 (IV)  &  9.95  &  9.92  &  9.57  &  9.05  &   8.54  &  3  &  0.13  &     &     &      &  2 0 0 0  &  1.62  \\  
8322  &  952  &  1&  16 04 36.2  &  -51 38 36  &  143751  &   B3 II/III  &  10.23  &  9.72  &  8.56  &  8.45  &   8.36  &  0  &  0.28  &     &     &      &  4 0 0 0  &  2.48  \\  
8323  &  688  &  1&  16 12 43.0  &  -51 35 54  &  145300  &   F0 III  &  10.91  &  10.44  &  9.46  &  9.29  &   9.21  &  0  &  0.21  &     &     &      &  4 0 0 0  &  3.00  \\  
8319  &  698  &  1&  16 18 39.4  &  -49 24 49  &  146444  &   B2 VNE  &  7.67  &  7.60  &  7.15  &  7.09  &   6.93  &  0  &  0.29  &     &     &      &  4 0 0 0  &  1.04  \\  
8715  &  1647  &  1&  16 19 14.2  &  -54 57 42  &  146463  &   B3 VNNE  &  8.05  &  8.08  &  7.92  &  7.90  &   7.71  &  0  &  0.15  &     &     &      &  3 0 0 0  &  1.11  \\  
8324  &  1544  &  1&  16 19 42.8  &  -51 02 04  &  146630  &   K2 III  &  8.57  &  6.76  &  3.96  &  3.21  &   3.06  &  0  &  3.53  &  2.11  &  3.76  &   12.00  &  4 4 4 4  &  0.39$^{21}$  \\  
8333  &  1004  &  1&  16 34 43.5  &  -49 33 09  &  330950  &   B1VNE  &  10.05  &  9.78  &  8.66  &  8.56  &   8.50  &  0  &  0.14  &     &  1.01  &      &  2 0 1 0  &  1.85  \\  
7862  &  115  &  1&  16 35 48.7  &  -42 07 22  &  149313  &   B2 IA/BE  &  9.95  &  9.48  &  7.95  &  7.72  &   7.48  &  0  &  0.17  &     &     &      &  3 0 0 0  &  0.97  \\  
8333  &  1429  &  1&  16 36 11.8  &  -49 15 47  &  149298  &   B2 (II)  &  10.25  &  9.77  &  8.12  &  7.91  &   7.65  &  0  &  0.16  &     &     &      &  2 0 0 0  &  1.06  \\  
8329  &  1025  &  1&  16 38 57.7  &  -47 24 02  &  149742  &   K2 III  &  10.23  &  8.72  &  6.22  &  5.60  &   5.43  &  0  &  0.99  &  0.94  &  0.48  &      &  4 1 1 0  &  0.87  \\  
8338  &  2080  &  1&  16 47 51.4  &  -51 46 04  &  151083  &   B2 VN  &  9.25  &  9.08  &  8.21  &  8.04  &   7.76  &  0  &  0.15  &     &     &      &  2 0 0 0  &  1.15  \\  
7868  &  448  &  1&  16 51 00.0  &  -37 30 52  &  151771  &   B8 II/III  &  6.33  &  6.23  &  7.79  &  7.72  &   7.66  &  1  &  0.32  &     &     &      &  4 0 0 0  &  1.92  \\  
8335  &  1898  &  1&  17 00 28.7  &  -49 15 14  &  153222  &   B1 IB/IIIE  &  9.17  &  8.90  &  8.38  &  8.30  &   8.18  &  0  &  0.15  &     &     &      &  2 0 0 0  &  1.61  \\  
7372  &  375  &  1&  17 05 52.8  &  -36 35 17  &  154243  &   B3 VNNE  &  8.28  &  8.08  &  7.29  &  7.14  &   6.92  &  0  &  0.29  &     &     &   2.42  &  4 0 0 1  &  1.02  \\  
7878  &  1149  &  1&  17 16 17.6  &  -42 20 20  &  155896  &   B2/3 (V)NNE  &  7.03  &  6.96  &  6.20  &  6.12  &   5.99  &  0  &  0.63  &     &     &      &  4 0 0 0  &  0.97  \\  
7374  &  641  &  1&  17 18 38.3  &  -36 05 13  &  156369  &   A2 V  &  8.94  &  8.32  &  6.76  &  6.61  &   6.44  &  0  &  0.37  &     &     &      &  4 0 0 0  &  0.83  \\  
7874  &  809  &  1&  17 19 16.8  &  -39 48 25  &  156409  &   B2 (II)NE  &  9.18  &  8.78  &  7.94  &  7.80  &   7.59  &  0  &  0.18  &     &     &      &  3 0 0 0  &  1.20  \\  
7384  &  16  &  1&  17 37 30.9  &  -35 19 59  &  159684  &   B2 VNE  &  8.63  &  8.24  &  6.80  &  6.60  &   6.34  &  0  &  0.53  &     &     &      &  4 0 0 0  &  1.07  \\  
6840  &  1414  &  1&  17 48 35.5  &  -29 57 28  &  316341  &   B0,5V(PE)?  &  9.74  &  9.15  &  7.60  &  7.34  &   6.96  &  1  &  0.26  &     &     &      &  4 0 0 0  &  0.85  \\  
7377  &  827  &  1&  17 49 08.6  &  -31 17 16  &  161839  &   B5/7 II/III  &  9.77  &  9.64  &  7.63  &  6.58  &   6.35  &  1  &  1.33  &  0.89  &  1.04  &      &  4 1 2 0  &  2.09  \\  
6841  &  1403  &  1&  17 59 47.6  &  -23 48 58  &  163955  &   B9 V  &  4.70  &  4.72  &  4.84  &  4.61  &   4.47  &  0  &  1.29  &  0.72  &  0.68  &      &  4 1 2 0  &  0.78$^{J}$  \\  
7382  &  1514  &  1&  17 59 56.4  &  -33 24 29  &  163868  &   B2/3 (V)NE  &  7.30  &  7.35  &  7.14  &  7.06  &   6.88  &  0  &  0.31  &     &     &      &  3 0 0 0  &  1.07  \\  
6846  &  437  &  1&  18 04 39.6  &  -26 01 07  &  164950  &   B3/5 II  &  9.43  &  9.13  &  8.07  &  8.00  &   7.77  &  0  &  0.13  &     &     &      &  2 0 0 0  &  1.01  \\  
6263  &  2940  &  1&  18 04 43.2  &  -20 56 44  &  165014  &   F2 V  &  10.11  &  9.33  &  7.42  &  7.04  &   6.77  &  0  &  1.08  &  0.78  &  0.76  &      &  4 1 1 0  &  2.27  \\  
6259  &  2481  &  1&  18 05 58.8  &  -19 57 13  &  165285  &   B1/2 (I)NN(E)  &  8.89  &  8.45  &  7.02  &  6.78  &   6.48  &  0  &  0.44  &     &     &      &  4 0 0 0  &  1.00  \\  
6263  &  2328  &  1&  18 07 11.4  &  -21 26 38  &  165516  &   B1/2 IB  &  6.32  &  6.26  &  6.00  &  5.99  &   5.90  &  0  &  0.60  &  0.88  &     &      &  4 1 0 0  &  0.83  \\  
6272  &  2074  &  1&  18 08 27.1  &  -19 52 07  &  165783  &   B3/5 IB  &  8.69  &  8.37  &  7.45  &  7.28  &   7.09  &  0  &  0.22  &     &     &      &  4 0 0 0  &  0.92  \\  
6268  &  2490  &  1&  18 10 18.3  &  -18 11 41  &  166188  &   B3 (II)E  &  9.31  &  8.98  &  7.68  &  7.52  &   7.28  &  0  &  0.23  &     &     &      &  4 0 0 0  &  1.09  \\  
6268  &  212  &  1&  18 10 38.8  &  -17 44 02  &  166288  &   B5 IB  &  10.38  &  10.06  &  9.06  &  8.99  &   8.92  &  0  &  0.28  &     &     &      &  4 0 0 0  &  3.03  \\  
6268  &  620  &  1&  18 10 42.1  &  -17 55 05  &  166289  &   K1/2 (III)  &  11.17  &  9.31  &  6.35  &  5.64  &   5.45  &  0  &  1.45  &  1.11  &  1.13  &   1.89  &  4 2 1 1  &  1.30  \\  
6268  &  637  &  1&  18 12 33.9  &  -18 23 51  &  166691  &   B8 II  &  8.23  &  8.20  &  7.92  &  7.89  &   7.43  &  0  &  1.23  &  1.61  &  1.39  &      &  4 3 3 0  &  3.00  \\  
5689  &  949  &  1&  18 19 04.9  &  -13 48 20  &     &   B 1.5V  &  11.12  &  10.69  &  9.58  &  9.45  &   9.31  &  1  &  0.44  &  1.05  &  1.75  &   2.44  &  4 1 3 2  &  3.88  \\  
5689  &  815  &  1&  18 19 05.6  &  -13 54 50  &     &   B1V  &  9.73  &  9.50  &  8.71  &  8.65  &   8.61  &  0  &  0.21  &  0.90  &  1.17  &   4.21  &  4 1 3 4  &  2.43  \\  
6269  &  1768  &  1&  18 19 27.2  &  -18 13 10  &  168229  &   B2/3 IB/IIE  &  8.86  &  8.68  &  7.90  &  7.76  &   7.50  &  0  &  0.17  &     &     &      &  2 0 0 0  &  0.98  \\  
6265  &  875  &  1&  18 20 22.2  &  -15 48 29  &  168446  &   G8 V  &  10.76  &  9.91  &  8.21  &  7.86  &   7.77  &  0  &  0.26  &     &     &      &  4 0 0 0  &  1.79  \\  
6269  &  980  &  1&  18 22 50.3  &  -17 26 48  &  313240  &   B2 IV(NN)  &  9.51  &  9.34  &  8.57  &  8.44  &   8.25  &  1  &  0.12  &     &     &      &  2 0 0 0  &  1.36  \\  
6274  &  832  &  1&  18 27 12.0  &  -18 57 13  &  169805  &   B2 VNE  &  8.18  &  8.04  &  7.42  &  7.30  &   7.05  &  0  &  0.30  &     &     &      &  4 0 0 0  &  1.16  \\  
5124  &  2597  &  1&  18 36 07.8  &  -06 44 31  &  171610  &   K2 III  &  8.67  &  7.09  &  4.77  &  3.66  &   4.42  &  0  &  1.75  &  0.83  &  0.55  &   9.55  &  4 4 4 1  &  0.86  \\  
5700  &  654  &  1&  18 39 39.8  &  -11 52 42  &  172252  &   B0 V:e  &  10.12  &  9.63  &  7.93  &  7.74  &   7.49  &  0  &  0.17  &     &     &      &  3 0 0 0  &  0.98  \\  
5125  &  325  &  1&  18 44 33.3  &  -07 06 38  &  173219  &   B0 Iae  &  8.04  &  7.90  &  7.14  &  7.05  &   6.80  &  0  &  0.32  &     &     &      &  4 0 0 0  &  0.99  \\  
1044  &  195  &  1&  19 04 08.0  &  +11 06 24  &  230579  &   B1,5:IV:NE  &  9.72  &  9.19  &  7.74  &  7.54  &   7.26  &  1  &  0.22  &     &     &      &  4 0 0 0  &  1.02  \\  
2139  &  1057  &  1&  19 42 50.9  &  +22 33 45  &  344800  &   B2V:NNE  &  10.42  &  10.05  &  8.90  &  8.78  &   8.49  &  0  &  0.10  &  1.13  &     &      &  2 1 0 0  &  1.39  \\  
2682  &  3762  &  1&  19 59 55.2  &  +37 02 34  &  189687  &   B3V  &  4.97  &  5.12  &  5.41  &  5.48  &   5.49  &  0  &  0.85  &     &     &      &  4 0 0 0  &  0.83  \\  
2679  &  717  &  1&  20 09 58.4  &  +35 29 45  &  228041  &   B0,5V:E  &  9.35  &  9.05  &  7.87  &  7.64  &   7.36  &  0  &  0.19  &     &     &      &  3 0 0 0  &  0.98  \\  
2683  &  1156  &  1&  20 13 33.0  &  +36 19 42  &  192445  &   B0,5III  &  7.03  &  7.08  &  7.13  &  7.08  &   6.91  &  0  &  0.32  &     &     &      &  4 0 0 0  &  1.13  \\  
2683  &  174  &  1&  20 13 50.3  &  +36 37 22  &  228438  &   B0(IV)  &  8.77  &  8.41  &  7.18  &  7.05  &   6.76  &  0  &  0.30  &     &     &      &  4 0 0 0  &  0.85  \\  
2686  &  693  &  1&  20 33 05.1  &  +31 39 25  &  195907  &   B 1.5V  &  7.82  &  7.79  &  7.46  &  7.43  &   7.28  &  0  &  0.19  &     &     &      &  2 0 0 0  &  0.96  \\  
3171  &  1117  &  1&  20 54 22.3  &  +40 42 10  &  199218  &   B 8 V:NN  &  6.59  &  6.69  &  6.69  &  6.73  &   6.70  &  1  &  0.26  &     &     &      &  4 0 0 0  &  0.77  \\  
3575  &  553  &  1&  20 57 59.4  &  +46 28 00  &     &   M2IA  &  11.66  &  8.48  &  2.77  &  1.74  &   1.29  &  0  &  113.22  &  119.72  &  80.39  &   86.49  &  4 4 4 4  &  1.76  \\  
3970  &  673  &  1&  21 24 30.3  &  +55 22 00  &  204116  &   B1VE  &  8.13  &  7.60  &  6.21  &  5.96  &   5.63  &  0  &  0.88  &     &  0.89  &      &  4 0 1 0  &  0.88  \\  
3194  &  2107  &  1&  21 25 02.4  &  +44 27 06  &     &   B1,5V:PNNE  &  9.47  &  8.94  &  7.43  &  7.21  &   6.94  &  0  &  0.26  &     &     &      &  4 0 0 0  &  0.91  \\  
3979  &  1275  &  1&  21 36 59.6  &  +58 08 24  &  239712  &   B 2 VNNE  &  9.00  &  8.59  &  7.54  &  7.48  &   7.28  &  0  &  0.23  &     &     &      &  2 0 0 0  &  1.12  \\  
3986  &  1670  &  1&  22 18 45.6  &  +56 07 33  &  211853  &   B0:I:+WR  &  9.44  &  9.06  &  7.78  &  7.60  &   7.41  &  0  &  0.18  &  1.18  &  1.09  &      &  3 1 1 0  &  1.00  \\  
4282  &  933  &  1&  22 56 42.6  &  +62 37 29  &  217061  &   B1V  &  9.51  &  8.88  &  7.26  &  7.13  &   7.03  &  0  &  0.64  &  1.85  &  2.71  &   5.70  &  4 2 4 4  &  2.03  \\  
4011  &  773  &  1&  23 20 34.3  &  +58 16 39  &  220116  &   B0,5VPE  &  9.26  &  8.68  &  7.59  &  7.50  &   7.36  &  0  &  0.17  &     &     &      &  2 0 0 0  &  0.90  \\  
4285  &  1070  &  1&  23 49 53.1  &  +62 12 50  &  223501  &   B 1 VNE  &  7.73  &  7.74  &  7.74  &  7.77  &   7.67  &  0  &  0.18  &     &     &      &  3 0 0 0  &  1.28  \\  

\label{finaltable}

\end{longtable}
\end{landscape}

\end{onecolumn}
\begin{twocolumn}

\appendix

\section{Other MSX catalogues}
\label{app}
The MSX satellite observed a number of regions in addition to the
galactic plane and the IRAS gaps. These were generally regions of high
source density that were labelled as confused by IRAS and included the
Magellanic Clouds, several star forming regions and a number of
extended objects such as galaxies. We have undertaken a search for
excess stars in the IRAS gaps and all of the above regions except the
galaxies. The results of these searches are discussed below.

A positional correlation of the non-plane ($\mid b \mid>6$\degr) MSX
catalogue sources with the Tycho 2 star catalogue found 5898 (68\%)
MSX sources within 6\arcsec of a Tycho 2 star. The non-plane MSX
catalogue contains both the IRAS gap (section \ref{irasgap}) and the
Large Magellanic Cloud (LMC) surveys.  We discuss the LMC stars
separately in section \ref{magellanic} and define these to be within a
14\degr square (272\degr,-26\degr to 286\degr,-40\degr) around the
LMC's galactic coordinate position.  Of the non-plane sample 753
(12\%) were within 45\arcsec of an IRAS source. A small fraction
(20\%) of these IRAS associations are LMC stars.

\subsection{IRAS gaps}
\label{irasgap}

5364 MSX-Tycho 2 IRAS gap stars are identified.  66 \% (3522) of these
have spectral type information and 5215 have 2MASS counterparts.  A
near-mid IR colour diagram is shown in Figure \ref{sfr}.
Although much less reddened, the $H$$-$$K$ $K$$-$$[8]$ colours are similar to
the galactic plane sample. The same excess determination process was
applied to the IRAS gaps, and we identify 95 (87 warm and 8 hot) IR
excess stars in this region.  22 of these excess stars were found to
be within 45\arcsec of a IRAS source; these were found to be at the
edges of the IRAS gap regions.
 
The star with large excess is the variable star ST Pup, the other
stars with excesses fall neatly into the groups we have previously
discussed for the galactic plane. These stars are published in Table
3.

\subsection{Magellanic Clouds}
\label{magellanic}

The LMC was found to contain 534 MSX-Tycho 2 sources, of these 523
have 2MASS counterparts and 333 have spectral type information. A
$H$$-$$K$ $K$$-$$[8]$ colour diagram of the sample is shown in Figure
\ref{sfr}. These stars are mainly G,K and M giant type stars. However
a comparison of MSX and 2MASS observations of the LMC by \citet{egan}
indicated that the spatial distribution of these stars are more likely
galactic foreground objects rather than LMC stars. We identify 24 warm
excess stars, 10 of which were within 45\arcsec of a IRAS
source. The LMC infrared excess stars can be found in Table 4.

The excess stars are nearly all supergiants, the objects with
$H$$-$$K>0.70$ are well known Be supergiants, their extreme $H$$-$$K$
colours are a result of strong $K$ band excess from the stars hot dust
disk. Interestingly the colours of the Be supergiants overlap with the
Herbig stars observed in the galactic plane (see
Fig.\ref{nearcolour}). The stars with blue $H$$-$$K<0.2$ colours and
large excesses are young massive stars. The lower excess stars are
late type stars with circumstellar material and the well known
Luminous Blue Variable S Dor.

A correlation of the Small Magellanic Cloud (SMC) MSX mini-catalogue
(containing 243 sources) with Tycho 2, yielded 55 associations. 54 of
these stars have 2MASS counterparts and 42 have spectral types. A
$H$$-$$K$ $K$$-$$[8]$ colour diagram of the SMC sample is shown in
Figure \ref{sfr}. We identify 4 excess stars (4 warm). Of these none
were detected by IRAS. The excess stars found in the SMC were known or
suspected late type supergiants. The SMC infrared excess stars are
listed in Table 5.

\subsection{Star forming regions outside of the galactic plane}
\label{starform}

Here we discuss only the regions that are outside of the galactic
plane, and hence are not part of the galactic plane survey discussed
earlier. The following star forming regions were imaged during the MSX
mission (see \citealt{Kr} for full details):

The Orion Nebula (A \& B), the HII region S263, the IRAS loop
G159.6-18.5 associated with the Perseus Molecular Cloud, the Pleiades
star cluster and G300.2-16.8 an isolated cloud in Chamaeleon
associated with IRAS 11538-7855. We have searched the MSX Point Source
mini-catalogues associated with each of the regions for Tycho 2 and
2MASS counterparts.  The searches found 160 MSX-Tycho2-2MASS sources
in Orion, 41 in S263, 53 in G159.6-18.5, 51 in the Pleiades and 54 in
G300.2-16.8.

We show the $K$$-$$[8]$ $H$$-$$K$ colour diagram for these star
forming regions in Figure \ref{sfr}. The majority (78 \%) of the IR
excess stars in the star forming regions are found in the Orion
Nebula. In the star forming regions we identify 51 excess stars (46
warm 4 hot 1 cool). Of theses sources approximately 80 \% were within
45\arcsec of an IRAS point source. The excess stars tend to split into
the Herbig stars with $H$$-$$K>0.5$ and variable stars of Orion type
$H$$-$$K<0.5$. The infrared excess stars found in the star forming
regions are in Table 6.

\begin{figure*}
\includegraphics[scale=0.95]{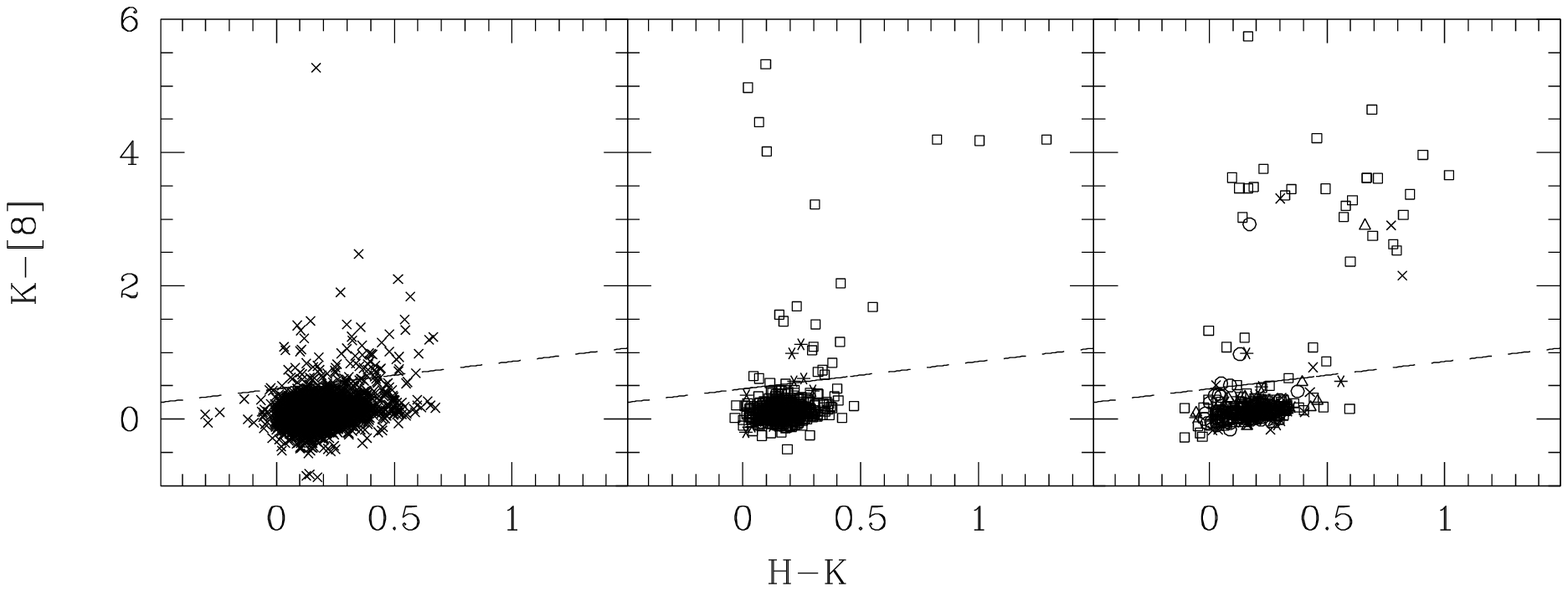}
\caption{Near-Mid IR colour diagram of the IRAS Gaps (5215 stars)
(left),the LMC $\Box$ (523 sources) SMC $\ast$ (54 sources)
(centre) and the non-galactic plane star forming regions observed
by MSX (right). In the right hand panel, the different regions
are indicated by the following symbols: $\times$ G159.6-18.5,
$\triangle$ G300.2-16.8, $\Box$ Orion Nebula (A and B), $\circ$
Pleiades and a $\ast$ for S263. The excess cutoff is shown in all the
diagrams by a broken line.}
\label{sfr}
\end{figure*}

\end{twocolumn}
\end{document}